# Sustainable Green Networking: Exploiting Degrees of Freedom towards Energy-Efficient 5G Systems

Miao Yao, Munawwar Sohul, Xiaofu Ma, Vuk Marojevic, Jeffrey H. Reed

Bradley Dept. Electrical and Computer Engineering, Wireless@Virginia Tech

miaoyao@vt.edu

## Abstract

The carbon footprint concern in the development and deployment of 5G new radio systems has drawn the attention to several stakeholders. In this article, we analyze the critical power consuming component of all candidate 5G system architectures—the power amplifier (PA)—and propose PA-centric resource management solutions for green 5G communications. We discuss the impact of ongoing trends in cellular communications on sustainable green networking and analyze two communications architectures that allow exploiting the extra degrees-of-freedom (DoF) from multi-antenna and massive antenna deployments: small cells/distributed antenna network and massive MIMO. For small cell systems with a moderate number of antennas, we propose a peak to average power ratio-aware resource allocation scheme for joint orthogonal frequency and space division multiple access. For massive MIMO systems, we develop a highly parallel recurrent neural network for energy-efficient precoding. Simulation results for representative 5G deployment scenarios demonstrate an energy efficiency improvement of one order of magnitude or higher with respect to current state-of-the-art solutions.

## 1. Introduction

The global climate change has emerged as a critical issue over the last decades. The increasing popularity of wireless communication networks, has resulted in information and communication technology (ICT) becoming a non-negligible contributor to the overall carbon footprint [1]. With the assumption of relatively constant energy efficiency, the increasing number of base stations (BSs) and remote radio heads (RRHs) leads to higher operating expenditure (OPEX) mainly because of the higher energy consumption [2]. This growth can be attributed not only to the increase in the number of smart devices in emerging economies, but also to the growth of shared multimedia data and online games. The wireless industry needs significant improvements in the energy efficiency of BSs and other network infrastructure to compensate for the increased energy demands from the network growth [3] [4] [5]. Therefore, designing energy-efficient communication systems has become a critical issue for 5G, which promises massive deployment of smart devices served new infrastructure elements. The International Mobile Telecommunications' (IMT) view of the next generation cellular system indicates the expectation of 100x improvement in network energy efficiency by 2020 [1]. Concepts such as "sustainable green communications" have recently emerged and describe the common trend toward energy-efficient wireless communications systems.

It is most likely the shift from 4G long term evolution (LTE) to 5G new radio (NR) will be the first wireless standard migration without waveform change [6], although alternative multicarrier waveforms such as filter bank multicarrier (FBMC) and universal filtered multicarrier (UFMC) have also been proposed as promising 5G waveform candidates. As a typical multicarrier air interface, orthogonal frequency division multiplexing (OFDM) has been widely adopted as the air interface of various wireless communication systems such as 4G LTE and IEEE 802.11 family. Its success is associated with many advantages including robustness against multipath fading, high spectral efficiency, and the support for MIMO and orthogonal frequency division multiple access (OFDMA). All the aforementioned multicarrier waveforms suffer from a high peak to average power ratio (PAPR) to different extents at the transmitter as a result of the constructive addition of modulation symbols that are simultaneously carried over several narrowband subcarriers. The high peaks lead to signal excursions into nonlinear region of the high power amplifier (PA) and, thereby, to nonlinear signal distortion and spectral spreading. High input power back-off is required at the PA to keep the peaks of a multicarrier signal within the amplifier's linear region. In cellular networks, the high back-off requirement causes low energy efficiency at the BS, higher cost PAs, and increases the utility and cooling costs for the operator. As a result, approximately 60% of the total power consumption of a 4G macrocell BS is attributed to the PAs [7]. Hence, the PA is the most critical component to consider when targeting energy-efficient cellular network design and operation.

One of the most promising features of 5G that can bridge the network energy efficiency gap is large-scale antenna deployment, both as distributed antenna networks (DANs) and massive MIMO systems. We identify and examine the following key challenges in building PA-centric green and sustainable 5G communications and networking systems:

- Can a cloud radio access network (C-RAN) provide additional probability for energy efficiency improvement of PA-centric multiple antenna system?
- Should we apply different PA-centric green communication strategies for moderate and large scale antenna systems?
- Is the application of statistical or instantaneous PAPR-aware scheme related with the number of antennas?

Multiuser resource allocation for OFDMA systems has drawn the attention of many researchers. The authors of [8] introduce algorithms that allocate power to each subcarrier as a function of the channel state information (CSI) to provide optimal resource allocation. Growing economic pressures and environmental awareness have shifted the attention of designers towards energy efficiency. The fundamental tradeoffs between spectral efficiency and energy efficiency in resource allocation are discussed in [9]. On the other hand, there has been recent research to enhance the system energy efficiency by using multiple collocated or distributed antennas [2] [10] [11]. He et al. [2] proposes a suboptimal resource allocation algorithm to maximize the energy efficiency while considering the proportional fairness for different mobile stations (MSs) and discusses the tradeoff between energy efficiency and spectral efficiency for DAN. Ng et al. [11] propose an energy-efficient resource allocation algorithm with a large number of collocated antennas. However, these methods neglect the effect of power allocation on the PAPR but, rather, assume constant and equally efficient PAs. In other words, the interrelation among PAPR, drain efficiency and subcarrier resource allocation is neglected.

DAN is considered a promising technology for future wireless systems [2]. In a traditional DAN architecture, the remote radio heads (RRHs) are geographically distributed and connected to the central unit of the system. The multi-antenna subcarrier allocation provides numerous degrees-of-freedom (DoF) for jointly optimizing channel capacity and PAPR at each RRH. For this reason, the scheme proposed in the Case Study 1 provides capacity improvement using DAN over a single antenna OFDMA system. Massive

MIMO is also widely perceived as a leading candidate technology for 5G [12]. The high number of BS antennas requires a large number of PAs, one per antenna. Noticed the tradeoff between linearity and efficiency in Figure 1, highly linear and low efficiency PAs in a 5G BS would drive the CAPEX cost to infeasible figures. It is therefore critical for 5G cellular networks to use inexpensive PAs. Massive MIMO systems have the potential to reduce the instantaneous PAPR by means of the precoding matrix, which is used for suppressing multiuser interference (MUI). The main idea behind the Case Study 2 is to jointly perform multiuser MIMO (MU-MIMO) precoding and PAPR reduction with a recurrent neural network (RNN). This network exploits the numerous DoF in a massive MIMO systems. The analogy between a neural network and massive MIMO system is used to inspire a PAPR reduction technique that lends itself to parallel implementation.

The rest of this paper is organized as follows. Section 2 introduces the definition of energy efficiency and the impact of PA efficiency to energy-efficient communication systems. Section 3 discusses the trends of small cell and massive antenna in energy-efficient communication systems. Section 4 presents a case study of the architecture, analysis, and simulation results of distributed antenna network in C-RAN. Section 5 presents a case study of energy-efficient massive MIMO system. Section 6 concludes the paper.

## 2. Energy Efficiency and the Effect of PA Efficiency

The cellular network is composed of a variety of components in which the BS is the highest power consumer. Energy-efficient wideband wireless communications systems are very sensitive to nonlinear distortions caused by the radio frequency (RF) chain and the high power amplifier (HPA), in particular. Energy efficiency metrics have been proposed to evaluate the performance of the wireless network at different levels [13]. The energy efficiency at the BS level reflects the achievable data rate scaled by the system power consumption and can be expressed as

$$\varepsilon_{EE} = \frac{R(\boldsymbol{p})}{\sum_m \frac{1}{\eta_m} \sum_n p_{m,n} + P_r}, \tag{1}$$

where $R(\boldsymbol{p})$ represents the sum data rate and $\boldsymbol{p}$ the power allocation matrix, $\boldsymbol{p} = [p_{m,n}]_{M \times N}$. The first part in the denominator represents the power consumption of the PAs, where $\eta_m$ captures the PA efficiency (PAE) of the HPA at the $m$th antenna, which is a function the power allocation and the PAPR as illustrated in Figure 1(a) and captured by

$$\eta_m = c_1 + c_2 \frac{2\pi \Delta f^2 (P_{T,m} \sum n^2 p_{m,n} - (\sum n p_{m,n})^2)}{P_{T,m}^3}. \tag{2}$$

Symbols $c_1$ and $c_2$ are constants, $\Delta f$ is the subcarrier spacing and $P_{T,m}$ the transmit power constraint of $m$th antenna. Note that there are two fractions in (1) for energy efficiency calculation (one for data rate over power consumption, one for allocated power over PA efficiency), a simple expansion of traditional fractional programming [14] is necessary to find the optimal energy efficiency. The power consumption of the remaining part of the transmission system $P_r$ in (1) includes the power supply, baseband unit (BBU) for analog and digital signal processing and air conditioning.

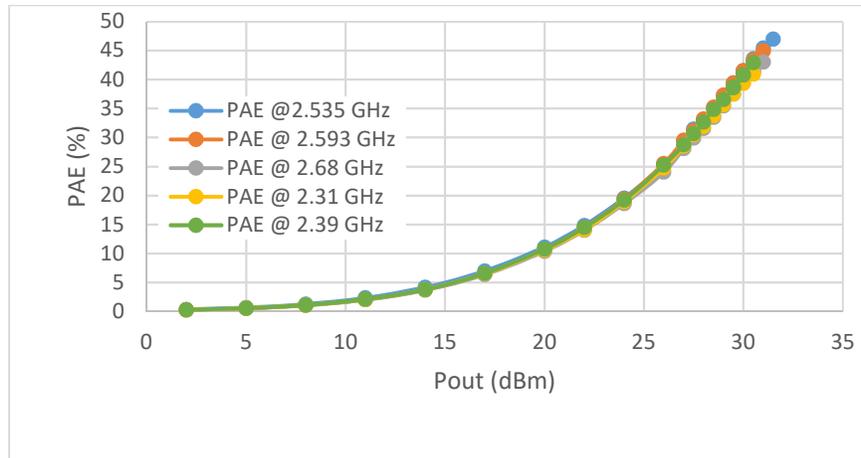

(a)

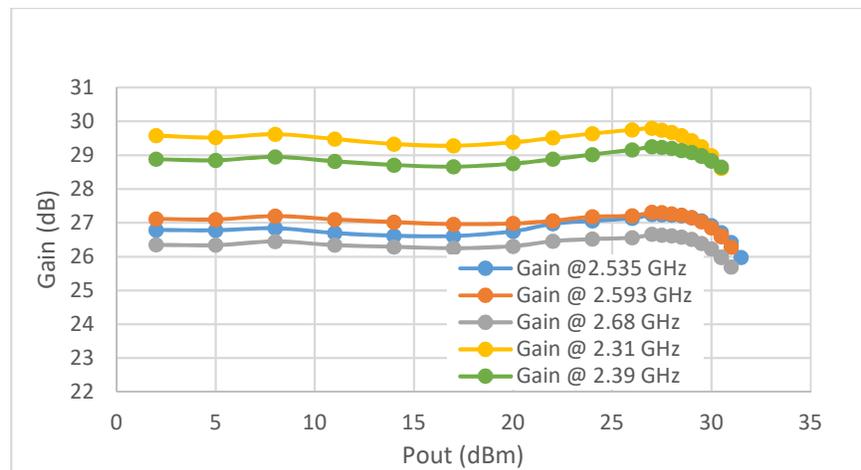

(b)

Figure 1: PA efficiency (PAE) vs Pout for LTE power amplifier; (a) PA gain vs Pout for LTE power amplifier at different operating frequencies (b) (Graphs obtained from real measurements for 20 MHz bandwidth).

The worst-case PAPR is proportional to the number of active subcarriers in the OFDM signal because the peak power is obtained as the constructive addition of the subcarriers' modulation symbols. The PA must be capable of accommodating the dynamic range of the signal determined by the PAPR. Therefore, the input power back-off in dB is defined for a PA by the difference between the operating point, which is usually the signal average power, and the peak power, which corresponds to the PAPR.

As shown in our measurement of Figure 1(b), the gain of a PA corresponds to the ratio between its output and input powers and remains relatively constant in the linear region, beyond which the gain starts to decrease due to the saturation. According to our measurement in Figure 1(a), the PA efficiency (PAE) increases with the transmitted power, where the saturation point of the PA delivers the highest operating efficiency. The operating point of a PA must be shifted to the lower power region to reduce the nonlinear distortion that occurs when reaching saturation. Typically, the PAE is reduced when reducing the average output power as determined by the appropriate input back-off value. For input signals with a large PAPR, the input back-off must be accordingly large to keep the peak power below the saturation level and make the PA working in its linear region. A signal with a higher PAPR also calls for more expensive PA with wider linear to maintain a reasonable average output power. The substantial power consumption of the PA leads to low energy efficiency of the entire BS [15]. Conversely, if the PAPR could be reduced, less expensive PAs would be needed while operating at higher efficiency. As a result, PAPR-aware resource allocation helps operators reducing capital and operational expenditures (CAPEX and OPEX) of network deployment and operation.

## 3. Improving Energy Efficiency by Small Cell and Increasing the Number of Antennas

As one of the most promising efforts towards sustainable green networking, the small cells provides a higher average data rate and better energy efficiency than a macrocell by reducing the cell size and, hence, the propagation loss [16] [17]. In this manner, a group of small cells which are enabled and managed by a C-RAN provide an energy efficiency gain in several ways [18] [19]:

- Lower CAPEX because of less overprovisioning due to infrastructure sharing;
- Lower OPEX because of more efficient maintenance (e.g. cooling), management, integration and upgrades;
- Lower transmission powers on downlink and uplink;
- Higher data to signaling/control ratio due to increased granularity for resource management and less diversified user/traffic profile and
- Comparatively higher achievable data rates due to spatial frequency reuse.

The RRH/C-RAN architecture lends itself to more effective resource usage and naturally provides a better energy efficiency [20]. The research question arises how to further leverage the centralized baseband processing and distributed antenna system to improve the energy efficiency of the RF front end, largely determined by signal statistics and PA performance figures.

The diversity of the wireless channel can be leveraged by a multiple antennas system using MIMO technology to increase the system capacity and reliability of one or several wireless communications links [21]. It is well known that the DoFs increase with the number of antennas. A higher DoF, moreover, allows selecting a larger signal space for resource allocation or precoding to achieve a comparatively lower PAPR in each RF chain. In addition, the power consumption of each antenna's PA is significantly reduced since the power is spread across various separate amplifiers in each radio antenna.

The evolution of MIMO, massive MIMO, uses orders of magnitude more antennas (e.g., 100 or more) at each BS, enabling even higher gains in terms of energy efficiency. Because of the large number of PAs to support massive MIMO in next generation cellular communication systems, inexpensive PAs need to be deployed. As the number of antennas at a BS grows, the fast fading effects of the channels can be neglected, and the PAPR reduce to almost 0 dB. Massive MIMO, moreover, helps reducing the complexity of single-antenna receivers, where even simple linear signal processing, such as maximal ratio combing (MRC), can provide the desired performance.

## 4. Case Study 1: Distributed Antenna Network in a C-RAN Architecture

*A) Communications Architecture*

In the context of a small cell, we assume that $M$ distributed antennas/RRHs are connected to the centralized baseband processing units of a C-RAN [2]. Figure 2 shows $K$ users that are served by a set of distributed antennas, sharing $N$ subcarriers. The subcarriers are allocated on SDMA group basis, which means that each subcarrier is exclusive to one group. Furthermore, all RRHs are involved in the transmission to the mobile stations (MSs) and any subcarrier carries the same information symbol across all the RRHs involved in the transmission. The power is dynamically allocated to the subcarriers of the different RRHs according to the channel conditions [10] [22]. Statistical PAPR-aware approach is applied in this case study since we take care of statistical characteristic of transmit signal.

*B) Problem Formulation*

We use the energy efficiency (3) as the objective function of our optimization framework to jointly account for data rate and PAPR. As shown in Figure 1(a), it is not practical to assume that the PA efficiencies are equal across all RRHs and independent of the power allocation scheme. We rather consider that the PA efficiency depends on the PAPR which determines the power back-off and, hence, the subcarrier power allocation. The optimal solution for subcarrier and power allocation can be obtained by solving

$$\text{Maximize} \quad \varepsilon_{EE}, \tag{3}$$

$$\text{Subject to} \quad \sum_n p_{m,n} \leq P_{T,m}, \forall\, m,$$

$$p_{m,n} \geq 0, \forall\, m, n,$$

$$\sum_{k \in \pi_i} \omega_{k,n} = 1, \forall\, n, i,$$

$$\sum_{k \in \pi_i} \omega_{k,n} \in \{0,1\}, \forall\, n, k, i.$$

Parameter $\omega_{k,n}$ indicates whether the $n$th subcarrier is allocated to the $k$th MS, $\pi_i$ represents the set of MSs in $i$th SDMA group, and $P_{T,m}$ represents the power constraints of the $m$th RRH.

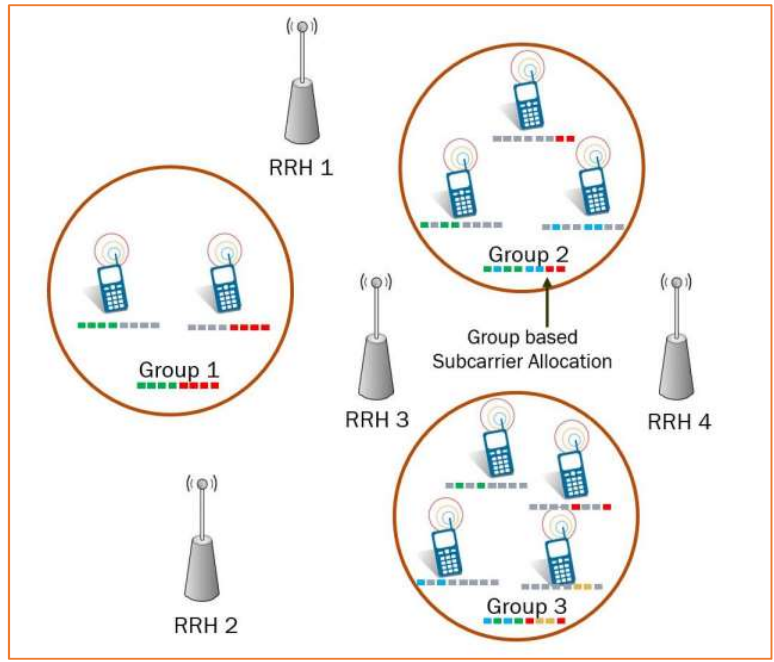

Figure 2: Structure of OFDMA/SDMA based DAN for small cell deployment with multiple distributed users (greedy based SDMA group is applied to allocate the MSs to different groups, subcarriers which represented by color blocks are assigned to different MSs within each group).

*C) Joint Channel Capacity and PAPR Optimization*

By combining OFDMA with space division multiple access (SDMA) the system can exploit the rich spatial diversity of DAN for further improving the spectral efficiency. The bandwidth of SDMA is reused by letting multiple users which are sufficiently spatially separated share the same subcarriers. A joint OFDMA/SDMA method with greedy grouping is proposed in this case study. The multi-antenna subcarrier allocation technique exploits the DoF at each RRH. In order to address this joint optimization problem, we derive the complementary cumulative distribution function (CCDF) and hence the PA efficiency for the OFDM waveforms with unequal power allocation (UPA) in (2), which generalizes the closed-form solution for the PAPR distribution of OFDM that has been developed for equal power allocation [23]. The closed-form expression for the PAPR distribution of an OFDM signal allows analyzing the interdependency between the PA efficiency and the resource allocation.

*D) Solution*

The joint OFDMA/SDMA communications system allocates subcarriers to MSs considering their spatial diversity. The MSs are therefore divided into several SDMA groups, the MSs with spatially correlated channels are placed in the same group and multiplexed on different resources, e.g. on different subcarriers. MSs are multiplexed by employing the SDMA scheme, e.g., with a transmit ZF filter, while reusing the same resources in frequency and time. The power is allocated to subcarriers of different RRHs afterwards. Suboptimal subcarrier allocation is carried out by considering only the sum capacity (without considering energy efficiency). Notice that there are two fractions in the energy efficiency (1), namely data rate over total power consumption and power allocation over PA efficiency. The subcarrier allocation is followed by the dual fractional programming to deal with the two fractions in (1) with a two-step approach [24].

The energy efficiency problem after application of two-step dual fractional programming is reduced to a simple water filling problem. Therefore, we introduce an iterative power allocation algorithm where the power is allocated to only one antenna in each iteration while the powers of the remaining antennas are assumed constant. Although the PAPR is considered during the resource allocation, the computing complexity of our DAN power allocation solution increases linearly with the number of distributed antennas $M$ and is thus suitable for massive deployments of RRHs.

*E) Numerical Results*

The simulations assume small cells within the service area of 2 kilometer radius. The distributed antenna network consists of 5-40 RRHs, serving 50 MSs using 128 subcarriers and QPSK. A lower number of RRHs helps to evaluate the performance of the PAPR reduction with limited degradation on channel capacity, whereas a high number of RRHs are considered to evaluate the performance of the energy efficiency optimization with practically unlimited DoF. The MSs are randomly distributed within the range of the small cell. The C-RAN's BBUs are located in the center of the service area. The RRHs are uniformly distributed. The channels between RRH and users are modeled as independent frequency selective channels.

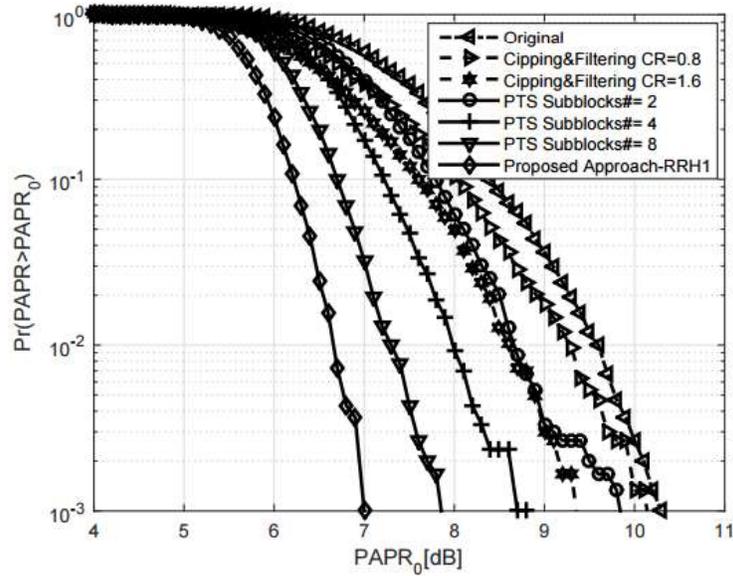

Figure 3: Performance comparison of different PAPR reduction schemes.

Figure 3 shows the performance in terms of PAPR reduction of the proposed approach along with state-of-the-art PAPR reduction schemes. The clipping rate (CR) is chosen to be 0.8 and 1.6 in the simulations for clipping and filtering scheme [25]. The number of subblocks, which determine the granularity of optimization, in partial transmit sequence (PTS) [26] are chosen to be 2, 4 and 8, respectively, and each partitioned subblock is multiplied by a corresponding complex phase factor. The graph shows that the proposed PAPR-aware energy efficiency optimization approach can provide consistent improvement when compared with traditional clipping and filtering or PTS approaches.

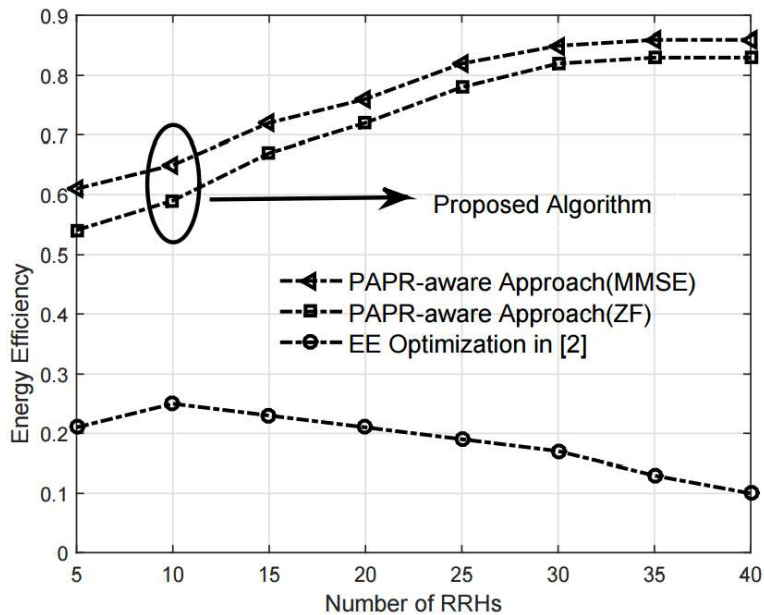

Figure 4: Energy efficiency over the number of RRHs for PAPR-aware energy efficiency optimization.

Figure 4 plots the energy efficiency over the number of RRHs for the proposed solution and the approach from [2], which treats the PA efficiency of different RRHs as equal and constant numbers irrespective of power allocation results. The energy efficiency gain of the proposed solution is almost one order of magnitude higher. The general trend shows that the energy efficiency of the proposed method increases with the number of RRHs, more steeply when the number of RRHs is relatively low. The PAPR-aware energy-efficient algorithm increases with the number of RRHs because of the higher DoF, which provides more space for optimization. These results show that the maximum energy efficiency is achieved for as many as 30 RRHs. We also observe that the energy efficiency of the approach proposed in [2] increases when the number of RRH is low, but decreases when increasing the number of RRHs beyond the optimal point of 10 RRHs in this case.

*F) Conclusions*

This case study has considered the overall data rate as well as the PAPR influence on the PA and, thus, the energy consumption for resource allocation to a DAN enabled by RRH/C-RAN infrastructure. Dual fractional programming is applied to derive the optimal power allocation method for DANs and an iterative solution is proposed to reduce the computational complexity. Simulation results have shown that both PAPR reduction and energy efficiency are improved significantly when compared with traditional energy-efficient resource allocation schemes. The expectation of PAPR-aware energy-efficient DAN structure is very promising in two aspects: 1) the performance of PAPR reduction enables the operator to equip the RRHs with low-cost PAs, which helps the operator to reduce the CAPEX of the DAN deployment and; 2) the OPEX of DAN in terms of energy consumption can be reduced by up to 90% especially with the large scale DAN.

## 5. Case Study 2: Massive MIMO in 5G

*A) Communications Architecture and System Model*

The second case study proposes a novel PAPR reduction method for energy-efficient massive MIMO operation. The premise is to exploit the DoF of the large number of antennas with minimal effect on data rate and multi-user interference (MUI). Our solution mimics an artificial neural network (ANN) to find the minimal dynamic range of the signal with an optimal precoding matrix. Consider a downlink massive MIMO-OFDM system which has $M_r$ single antenna users and one BS equipped with $N_t$ antennas, the number of subcarrier is $N_c$. The number of BS antennas is significantly larger than number of users, i.e. $N_t \gg M_r$ [27]. It has been proven that for the single carrier case, the large scale MIMO system yields signals with unit PAPR across the antennas as the number of antennas $N_t$ approaches infinity for a finite number of user $M_r$ [27]. When considering an OFDM waveform, the problem becomes more complicated since the constraints of the PAPR and precoding are in different domains: The PAPR depends on the spectral content of the signal at each individual antenna, whereas the precoding mechanism depends on the signal across multiple antennas. Therefore, the overall signal representation across different antennas and subcarriers can be formulated by the single equation

$$s = H \underbrace{\begin{bmatrix} I_{1,1} & \cdots & I_{1,N_t} \\ \vdots & \ddots & \vdots \\ I_{N_c,1} & \cdots & I_{N_c,N_t} \end{bmatrix}}_{\widehat{H}} \underbrace{\begin{pmatrix} \widehat{X_1} \\ \vdots \\ \widehat{X_{N_t}} \end{pmatrix}}_{\widehat{X}}. \quad (4)$$

Symbol **s** is the information symbol and **H** the channel state. The permutation matrix is comprised by the $N_t \times N_c$ matrix $I_{m,n}$ whose entries are 1 for $1 \leq m \leq N_c$ and $1 \leq n \leq N_t$, all other entries being 0. $\widehat{\mathbf{X}}$ represents the signals transmitted from the $N_t$ antennas. Instantaneous PAPR-aware approach is applied in this case study since we deal with transmit signal on symbol basis.

## B) Problem Formulation

The PAPR-aware downlink massive MIMO system is formulated as an optimization problem to determine the optimal signal and maximum allowed dynamic range of the transmitted signal. Assuming that the channel state information is available at the transmitter (CSIT), certain signal preprocessing algorithms such as linear precoding are applied at the BS to eliminate the MUI at the receivers. Zero forcing (ZF) precoding is commonly applied in massive MIMO because of its simplicity and good performance. The user information symbols at the BS are mapped to the appropriate transmit antenna so that the information received by each user has minimal interference from the other users' signals. In order to derive the PAPR of the transmit signal, the frequency domain signal $\widehat{\mathbf{X}}$ needs to be transformed into the time domain signal $\tilde{\mathbf{x}}$. Therefore, the constraint of perfect MUI remover across different antennas and subcarriers is derived as (5a) and the overall massive MIMO-OFDM downlink problem formulated as

$$\underset{\tilde{x}}{argmin} \quad [\tilde{\mathbf{x}}^H \odot \tilde{\mathbf{x}} - \mathbf{e}^T y]^+ + y, \tag{5}$$

$$\text{Subject to} \quad \mathbf{s} = \mathcal{H}\tilde{\mathbf{x}}, \tag{5a}$$

$$y \geq 0 \tag{5b}$$

Operator $\odot$ denotes element-wise multiplication, $[x]^+ = \max(x, 0)$, and $\mathbf{e} = [1,1,\cdots,1]$ is a $N_c N_t$-vector. The maximum dynamic range is defined as the RNN activation variable $y$, which also represents the crest factor of the PA. The solution space $(\tilde{\mathbf{x}}, y)^T \in U_0$ is the convex set where the solution exists. It is clear that the RNN activation variable $y$ in represents the minimum PA efficiency among all the transmit antennas. For the given optimal RNN variable, the proposed structure guarantees that the signal $\tilde{\mathbf{x}}$ at all transmit antennas will not be distorted. The similarity between the nonlinear function $[x]^+$ in (5) and the activation function in a neural network as shown in Figure 5(a) allow us to adopt neural network to solve the problem.

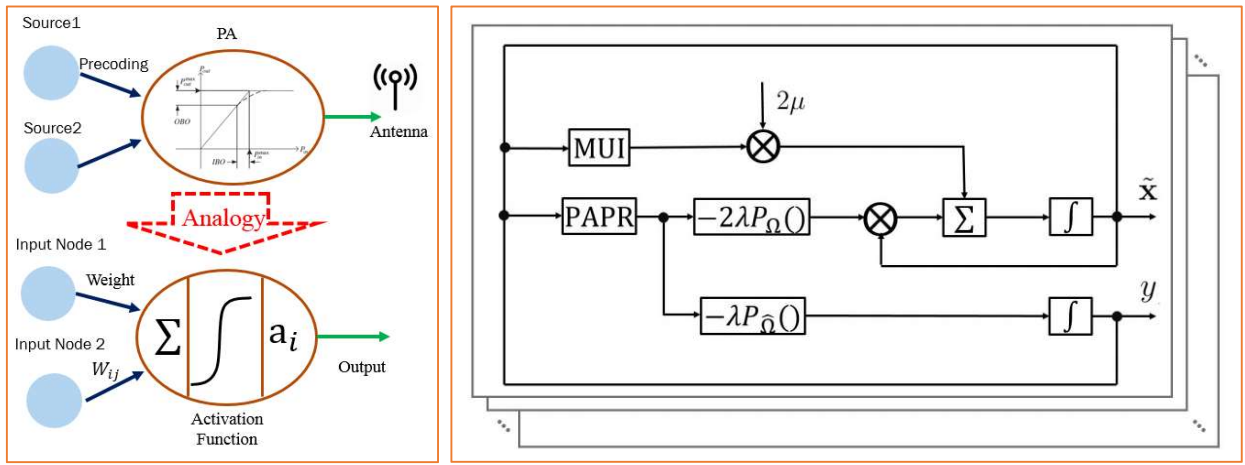

(a)                    (b)

Figure 5: Analogy between precoding & nonlinear PA structure and weighted sum and activation function neural network structure (a); Block diagram of the proposed recurrent neural network (b).

*C) Solution*

It has been demonstrated that the RNN is suitable for real-time implementation with finite and exponential convergence in various applications. In order to formulate the dynamic equation to derive the optimal precoding vector and RNN activation variable, we have to restate the optimization in (5) in Lagrangian form. First of all, let us relax the precoding constraint $s = \mathcal{H}\tilde{x}$ in (5a) to the Frobenius norm form $|s - \mathcal{H}\tilde{x}|_F^2 \leq \eta$, noting that the relaxation does not significantly degrade the BER performance for small values of $\eta$. With an auxiliary parameter $\lambda$, the scalar valued objective function of the RNN system is defined as the nonnegative Lagrangian function of the overall system

$$F(\tilde{x}, y) = \lambda([\tilde{x}^H \odot \tilde{x} - e^T y]^+ + y) + |s - \mathcal{H}\tilde{x}|_F^2. \tag{6}$$

With the nonnegative Lagrangian function defined in (6), the dynamic equations for solving (5) can be derived by taking the negative gradient

$$\frac{du}{dt} = -\mu \nabla F(u), \tag{7}$$

where $u = (\tilde{x}, y)^T \in U_0$, the nonnegative Lagrangian function $F(u)$ being defined in (6), $\nabla F(u)$ being the gradient of the objective function, and $\mu$ a positive scalar constant that is used to scale the convergence rate of the RNN. The block diagram of the proposed RNN is shown in Figure 5(b). The PAPR block represents the PAPR level of $(\tilde{x}^H \odot \tilde{x} - e^T y)$, and the MUI block the amount of MUI given by $\text{Re}(\mathcal{H}^H(\mathcal{H}\tilde{x} - s))$. Each layer in Figure 5(b) represent the RNN for each antenna.

The nonlinear activation function $P_\Omega(x_i)$ and deactivation function $P_{\hat{\Omega}}(x_i)$ are introduced to emulate the ideal cut-off characteristic beyond the 1-dB compression point in (8) and (9)

$$P_\Omega(x_i) = \begin{cases} 1, x_i \geq 0 \\ 0, x_i < 0 \end{cases} \tag{8}$$

$$P_{\hat{\Omega}}(x_i) = \begin{cases} 1, x_i < 0 \\ 0, x_i \geq 0 \end{cases} \tag{9}$$

The residual part of the signal after cut-off is back-propagated, and $\tilde{x}$ is trained by the signal residue to satisfy the precoding constraint in the meantime.

*D) Results*

The RNN-based large scale MU-MIMO-OFDM downlink system with 128 subcarriers, up to 250 antennas at the BS and 12 single-antenna MSs is chosen to evaluate the performance of the proposed PAPR-reduction scheme. The modulation is 16-QAM and the MSs are randomly distributed within the range of the small cell. The channels between each antenna and user are modeled as frequency selective channels and are independent from one another. The RNN-based scheme is simulated to derive the optimal transmitted signal and activation variable. The least squares (LS) method is used as a baseline, because it is one of the most prominent precoding methods for MIMO system. It generates the transmitted signal with minimum L2 norm while perfectly removing all the MUI.

The PAPR reduction performance of different PAPR-aware algorithm in time domain is shown in Figure 6. It is revealed in Figure 6 that the LS algorithm gives a solution with less power but the PAPR reduction performance is clearly worse than the other approaches. The fast iterative truncation algorithm (FITRA) [27] and the proposed RNN approach have similar performance in terms of PAPR reduction, while the proposed RNN approach has hardware-friendly parallel structure with less complexity.

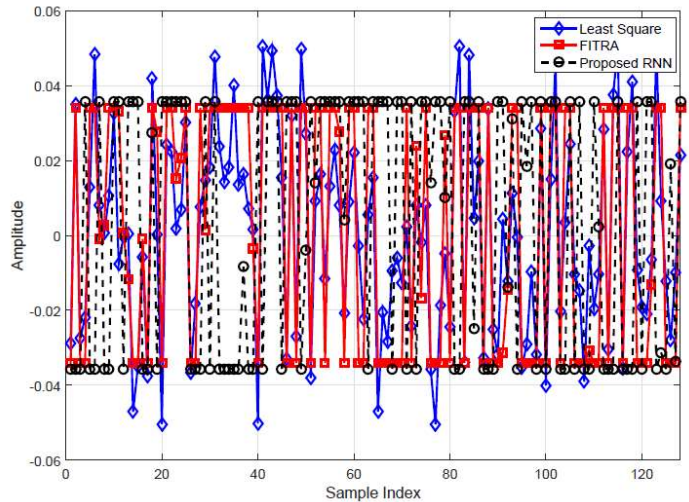

Figure 6: Transmit signal waveforms for different PAPR-aware precoding approaches in time domain

The relation between antenna configuration and energy efficiency performance is illustrated in Figure 7. The increasing number of antennas at the BS yields to better performance for the proposed RNN method, but to worse performance for the LS scheme.

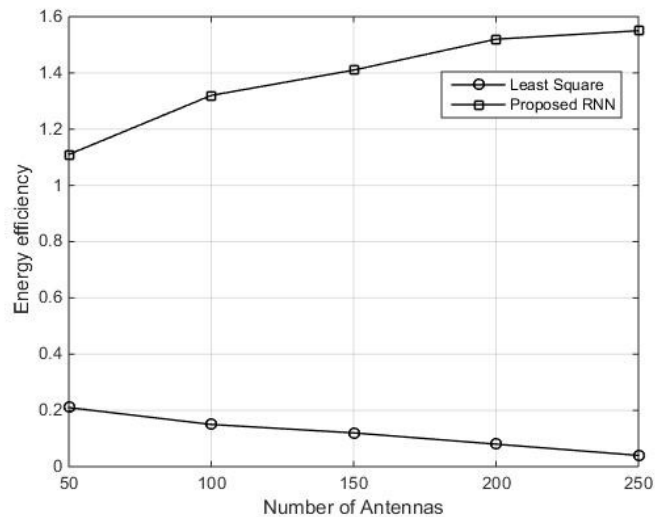

Figure 7: Energy efficiency of proposed RNN and least square precoding schemes as a function of the number of transmit antennas

*E) Conclusions*

The second case study has introduced a PAPR-aware massive MIMO-OFDM downlink system that is inspired by RNNs. It draws the analogy between the linear precoding with nonlinear PAs and massive-MIMO and the linear weighting of neurons with nonlinear activation functions of artificial neural networks. The proposed RNN provides a scalable, parallel processing solution which is especially suitable for tackling large-scale problems such as massive MU-MIMO. The near-constant envelope of the input signal is achieved by exploiting the extra DoF of the antenna array. The simulation results indicate that the DoF in PAPR-aware massive MIMO system can be exploited to improve the energy efficiency of the communications system. The proposed RNN scheme presents a simple processing structure, which is especially suitable for very-large scale integration (VLSI) implementation.

## 6. Conclusion

The high input power back-off power requirement for PAs causes low PA efficiency and hence low energy efficiency at the BS. Sustainable green networking systems need to be PAPR-aware. The trends for 5G NR, deploying RRH/CRAN or massive MIMO BSs, naturally provide more DoF that can be exploited to increase the net energy efficiency and reduce OPEX for the operator. This article has examined two candidate architectures for next generation cellular networks characterized by median-scale and large-scale antenna systems and developed the suitable resource allocation schemes. Numerical results highlight the vast potential for PAPR-aware green networking by utilizing the extra DoF in future multiple antenna communications systems.

## References

[1] ITU, "Report on the twenty-first meeting of working party 5D," 2015.

[2] C. He, G. Y. Li, F.-C. Zheng and X. You, "Energy-efficient resource allocation in OFDM systems with distributed antennas," *IEEE Transactions Vehicular Technology,* vol. 63, p. 1223–1231, 2014.

[3] H. Zhang, N. Liu, X. Chu, K. Long, A. H. Aghvami and V. C. & Leung, "Network Slicing Based 5G and Future Mobile Networks: Mobility, Resource Management, and Challenges," *IEEE Communications Magazine,* vol. 55, no. 8, pp. 138-145, 2017.

[4] M. M. Sohul, M. Yao, X. Ma, E. Y. Imana, V. Marojevic and J. H. Reed, "Next generation public safety networks: a spectrum sharing approach.," *IEEE Communications Magazine,* vol. 54, no. 3, pp. 30-36, 2016.

[5] M. M. Sohul, M. Yao, T. Yang and J. H. Reed, "Spectrum access system for the citizen broadband radio service," *IEEE Communications Magazine,* vol. 53, no. 7, pp. 18-25, 2015.

[6] Qualcomm White Paper, "Exploring 5G New Radio: Use Cases, Capabilities & Timeline," 2016.

[7] A. Gunther and et al., "How much energy is needed to run a wireless network?," *IEEE Wireless Communications,* vol. 18, no. 5, pp. 40-49, 2011.

[8] Z. Shen, J. G. Andrews and B. L. Evans, "Adaptive resource allocation in multiuser OFDM systems with proportional rate constraints," *IEEE transactions on wireless communications,* vol. 4, no. 6, pp. 2726-2737, 2005.

[9] C. Xiong, G. Y. Li, S. Zhang, Y. Chen and S. Xu, " Energy-efficient resource allocation in OFDMA networks," *IEEE Transactions on Communications,* vol. 60, no. 12, pp. 3767-3778, 2012.

[10] H. Zhang, S. Huang, C. Jiang, K. Long, V. C. Leung and H. V. Poor, "Energy efficient user association and power allocation in millimeter-wave-based ultra dense networks with energy harvesting base stations," *IEEE Journal on Selected Areas in Communications,* vol. 35, no. 9, pp. 1936-1947, 2017.

[11] D. W. Ng, E. S. Lo and R. Schober, "Energy-efficient resource allocation in OFDMA systems with large numbers of base station antennas," *IEEE Transactions on Wireless Communications,* vol. 11, no. 9, pp. 3292-3304, 2012.

[12] B. Han, S. Zhao, B. Yang, H. Zhang, P. Chen and F. Yang, "Historical PMI Based Multi-User Scheduling for FDD Massive MIMO Systems," in *Vehicular Technology Conference (VTC Spring), 2016 IEEE 83rd*, Nanjing, 2016.

[13] G. Auer and et al., "How much energy is needed to run a wireless network?," *IEEE Wireless Communications,* vol. 18, pp. 40-49, 2011.

[14] S. Schaible and J. Shi, "Fractional programming: the sum-of-ratios case," *Optimization Methods and Software,* vol. 18, no. 2, pp. 219-229, 2003.

[15] C. Han and et al., "Green radio: radio techniques to enable energy-efficient wireless networks," *IEEE Communications Magazine,* vol. 49, no. 6, pp. 46-54, 2011.

[16] H. Zhang, Y. Dong, J. Cheng, M. J. Hossain and V. C. Leung, "Fronthauling for 5G LTE-U ultra dense cloud small cell networks," *IEEE Wireless Communications,* vol. 23, no. 6, pp. 48-53, 2016.

[17] H. Zhang, C. Jiang, J. Cheng and V. C. Leung, "Cooperative interference mitigation and handover management for heterogeneous cloud small cell networks," *IEEE Wireless Communications,* vol. 22, no. 3, pp. 92-99, 2015.

[18] Z. Hasan and et. al., "Green cellular networks: A survey, some research issues and challenges," *IEEE Communications Surveys & Tutorials,* pp. 524-540, 2011.

[19] I. Gomez, V. Marojevic and A. Gelonch, "Resource management for software-defined radio clouds," *IEEE Micro,* vol. 1, no. 32, pp. 44-53, 2012.

[20] H. Zhang, Y. Qiu, X. Chu, K. Long and V. Leung, "Fog Radio Access Networks: Mobility Management, Interference Mitigation and Resource Optimization," *arXiv preprint arXiv:1707.06892.,* 2017.

[21] E. G. Larsson and e. al., "Massive MIMO for next generation wireless systems," *IEEE Communications Magazine,* vol. 52, no. 2, pp. 186-195, 2014.

[22] H. Zhang, Y. Nie, J. Cheng, V. C. Leung and A. Nallanathan, " Sensing time optimization and power control for energy efficient cognitive small cell with imperfect hybrid spectrum sensing," *IEEE Transactions on Wireless Communications,* vol. 16, no. 2, pp. 730-743, 2017.

[23] H. Ochiai and H. Imai, "On the distribution of the peak-to-average power ratio in OFDM signals," *IEEE Transactions on Communications,* vol. 49, no. 2, pp. 282-289, 2001.

[24] M. Yao, "Exploiting Spatial Degrees-of-Freedom for Energy-Efficient Next Generation Cellular Systems," *PhD Dissertation,* Virginia Tech.

[25] H. Ochiai and H. Imai, "Performance of the deliberate clipping with adaptive symbol selection for strictly band-limited OFDM systems," *IEEE Journal on Selected Areas in Communications,* vol. 18, no. 11, pp. 2270--2277, 2000.

[26] L. Yang, R.-S. Chen, Y.-M. Siu and K.-K. Soo, "PAPR reduction of an OFDM signal by use of PTS with low computational complexity," *IEEE Transactions on Broadcasting,* vol. 52, no. 1, pp. 83-86, 2006.

[27] C. Studer and E. G. Larsson, "PAR-aware large-scale multi-user MIMO-OFDM downlink," *IEEE Journal on Selected Areas in Communications,* vol. 31, pp. 303-313, 2013.